\documentclass[reprint,preprintnumbers,superscriptaddress, nofootinbib,amsmath,amssymb,aps,prd,floatfix]{revtex4-2}  
\pdfoutput=1
\usepackage{graphicx,amsmath,amssymb,amstext}
\usepackage{amssymb,amsbsy,amsfonts,amsthm,color}

\usepackage{epsfig}
\usepackage{graphicx}
\usepackage{subfigure}
\usepackage{hyperref}
\usepackage{cancel}
\usepackage{svg}
\usepackage{multirow}
\usepackage{parskip}
\usepackage{soul}
\usepackage[noabbrev]{cleveref}

\graphicspath{{figures/}}

\hypersetup{
  colorlinks   = true, 
  urlcolor     = blue, 
  linkcolor    = blue, 
  citecolor   = blue 
}

\usepackage{color}
\usepackage[dvipsnames]{xcolor}

\newcommand{\eqn}[1]{Eq.(\ref{#1})}

\newcommand{\mpl}{m_\mathrm{P}}
\newcommand{\dd}{\mathrm{d}}

\newcommand{\efolds}{\textit{e}-folds~}

\newcommand{\kcl}{Theoretical Particle Physics and Cosmology Group, Physics Department, King's College London, Strand, London WC2R 2LS, United Kingdom}
\newcommand{\ucsd}{Department of Physics, UC San Diego, 9500 Gilman Rd, La Jolla, CA, 92093, USA}
\newcommand{\upv}{Department of Physics, University of Basque Country, UPV/EHU, 48080, Bilbao, Spain}
\newcommand{\fzu}{CEICO, Institute of Physics of the Czech Academy of Sciences, Na Slovance 1999/2, 182 00, Prague 8, Czechia}

\begin{document}

\title{A critical value of the inflationary tensor-to-scalar ratio from inhomogeneous inflation} 
 \author{Panagiotis Giannadakis}
 \email{panagiotis.giannadakis@kcl.ac.uk}
 \affiliation{\kcl}
 
 \author{Matthew Elley}
 \email{elley@fzu.cz}
 \affiliation{\upv}
  \affiliation{\fzu}

 \author{Raphael Flauger}
 \email{flauger@ucsd.edu}
 \affiliation{\ucsd}
 
 \author{Eugene A. Lim}
 \email{eugene.a.lim@gmail.com}
 \affiliation{\kcl}

\begin{abstract}
We show that, for a given fixed value of the number of \efolds of the homogeneous solution, inflation succeeds with order unity inhomogeneities in the initial conditions above a characteristic value of the tensor-to-scalar ratio $r$. In practice, we work with an $\alpha$-attractor $T$-model and vary its characteristic scale $\mu$, keeping the initial inhomogeneities in both gradient and kinetic fields at ${\cal O}(1)$ of the inflationary energy scale. Under these conditions, and assuming 100 \efolds for the homogeneous solution, the requirement for 60 \efolds of inflation occurs at a critical characteristic scale $\mu_{\text{crit}} \approx 0.02\mpl$, corresponding to an $r_{\text{crit}} \approx 10^{-6}$.  Since increasing the amplitude of the inhomogeneities will make inflation less robust and hence require a higher characteristic scale in order for inflation to succeed, for a given number of \efolds achieved by the homogeneous solution $r_{\text{crit}}$ is a lower bound. 
\end{abstract}
\maketitle

\section{Introduction} \label{sect:intro}
The theory of cosmic inflation \cite{Guth:1980zm,Starobinsky:1980te,Linde:1981mu,Albrecht:1982wi}, which posits a period of accelerated expansion before the standard expansion of the Hot Big Bang, is the standard paradigm for the evolution of the very early universe. First introduced to solve several problems of Hot Big Bang cosmology, including the horizon and flatness problems, it offers a mechanism for generating an approximately scale-invariant spectrum of nearly Gaussian primordial quantum fluctuations, which later seed structure formation.

Of course, inflation only provides a solution to the horizon problem if it does not suffer from its own horizon problem. Since completely homogeneous initial conditions are arguably unique, it is natural to ask if inflation can begin given ``generic'' inhomogeneous initial conditions.  This issue, often called the ``initial conditions problem of inflation''\footnote{As opposed to the initial conditions problem of cosmology.} has been extensively studied both analytically \cite{Linde:1983gd, Linde:1984ir, Linde:1985ub,Gibbons:1977mu,Hawking:1981fz,Wald:1983ky,Starobinsky:1982mr,Barrow:1984zz,Albrecht:1984qt,Barrow:1985,Gibbons:1986xk,Jensen:1986nf,Hawking:1987bi,Penrose:1988mg,Muller:1989rp,Kitada:1991ih,Kitada:1992uh,Bruni:1994cv,Maleknejad:2012as,Gibbons:2006pa,Boucher:2011zj,Bruni:2001pc,Muller:1987hp,Barrow:1989wp,Bicak:1997ne,Capozziello:1998dq,Vachaspati:1998dy,Barrow:1987ia,Barrow:1986yf,Polyakov:2009nq,Marolf:2010nz,Tsamis:1992sx,Brandenberger:2002sk,Geshnizjani:2003cn,Marozzi:2012tp,Brandenberger:1990wu,Carroll:2010aj,Corichi:2010zp,Schiffrin:2012zf,Remmen:2013eja,Corichi:2013kua,Mukhanov:2014uwa,Remmen:2014mia,Berezhiani:2015ola,Kleban:2016sqm, Marsh:2018fsu,Finn:2018krt,Bloomfield:2019rbs, Azhar:2022yip, Albrecht:1985yf,Albrecht:1986pi,Easther:2014zga,Braden:2016tjn,Alho:2011zz,  Alho:2013vva,Brandenberger:1990xu} and numerically using full general relativity
\cite{Goldwirth:1989pr, Goldwirth:1989vz, Goldwirth:1991rj, Laguna:1991zs, KurkiSuonio:1987pq,KurkiSuonio:1993fg, East:2015ggf, Clough:2016ymm, Clough:2017efm, Aurrekoetxea:2019fhr, Joana:2020rxm, Corman:2022alv, Joana:2022pzo, Joana:2024ltg,Florio:2024pgm,Brady:2025zxp}. A comprehensive review of the latter can be found in the recent review \cite{Aurrekoetxea:2024ypv}.

While these numerical studies explore various initial conditions and inflationary models, a consistent trend is that models in which the part of the potential that supports inflation varies over a larger range in field space are more robust than models in which it varies over a smaller range~\cite{East:2015ggf,Clough:2016ymm}. More precisely, for a single scalar field minimally coupled to gravity, using the characterization introduced in \cite{Abazajian:2016yjj} which was motivated by \cite{Roest:2013fha,Creminelli:2014nqa}, we say that inflation driven by a potential with a high \emph{characteristic scale} is more robust than inflation driven by a potential with a low characteristic scale.  

Analytic arguments show that even for large and inhomogeneous kinetic and gradient energy densities, inflation will succeed as long as the scalar field values are confined to the slow-rolling plateau. This is also supported by numerical simulations. Large inhomogeneities collapse into black holes. Away from black holes, the solution rapidly becomes homogeneous (on the order of one or two $e$-folds), and the black holes are diluted by inflation. 

It appears that the primary cause of failure of inflation due to the presence of field inhomogeneities is when, in some region of space the scalar field falls into the minimum and subsequently drags the rest of the field down with it. As numerical simulations confirm, this happens more easily for potentials with a low characteristic scale than a high characteristic scale. It is then natural to ask whether we can identify a critical characteristic scale above which inflation is robust for generic initial conditions.

For single field inflation the characteristic scale directly maps to the tensor-to-scalar ratio $r$. So, identifying a critical characteristic scale would correspond to a critical value of $r$ that provides a target for experimental searches. Conversely, we may be able to gain some information about initial conditions of inflation from constraints on $r$.

We have referred to generic initial conditions multiple times, but the question of what constitutes ``generic'' in terms of initial conditions (see e.g. \cite{Albrecht:1984qt,Albrecht:1986pi,Berezhiani:2014kga,Berezhiani:2022gnv}) is a difficult problem. Without a UV complete description of inflation that allows us to define a measure on the space of initial conditions, we have to make some assumptions. 

From the perspective of effective field theory, if the single-field description becomes appropriate at a scale $\Lambda_{\text{UV}}^4\gg V_\text{inf}$, it is natural to consider 
\begin{equation}
    \rho_{\nabla,0} \approx \rho_{K,0} \gg V_{\text{inf}}\,. \label{eqn:grad}
\end{equation}
As we lower the amplitude of initial inhomogeneities, inflation is more likely to succeed; thus, the characteristic scale above which inflation succeeds decreases. Once the amplitude of initial inhomogeneities becomes so small that the potential dominates, the inhomogeneities become irrelevant to the question of whether inflation succeeds. To identify a critical value of the characteristic scale, it is then natural to consider a universe before inflation that approximately obeys ``equipartition'' i.e.
\begin{equation}
    \rho_{\nabla,0} \approx \rho_{K,0} \approx  V_{\text{inf}}~,\label{eqn:equipartition}
\end{equation}
where $\rho_{\nabla,0}$ and $\rho_{K,0}$ are the average gradient and kinetic energies, with $V_{\text{inf}}$ being the potential energy of inflation.

The critical value of the characteristic scale associated with~\eqref{eqn:grad} will be higher, so that the critical value we find is a lower bound on $r$ for a fixed number of \efolds achieved by the homogeneous solution.

For our numerical simulations, we focus on the $T$-model of the $\alpha$-attractor family models of inflation~\cite{Kallosh:2013hoa,Kallosh:2013yoa}. In this case, the characteristic scale $\mu$ can easily be specified and mapped directly to $r$. We consider initial conditions consistent with~\eqn{eqn:equipartition} and assume $100$~\efolds of inflation in the absence of inhomogeneities. We then vary $\mu$ to determine the critical value of the characteristic scale above which inflation is robust. We simulate using the numerical relativity code \textsc{grchombo}~\cite{Clough_2015,Andrade_2021,Radia_2022}. Due to computational limitations we cannot evolve the whole spacetime for the entire time inflation may last. Fortunately, it is not necessary to do so, as after several \efolds we can identify approximately homogeneous inflating patches on our grid. As long as the field in such regions is high enough on the plateau to result in more than 60 $e$-folds, and the size of the region is greater than or equal to the associated Hubble scale, we deem that the system has successfully inflated (despite potential failure in other regions). We found that there exists a critical $\mu_{\text{crit}} \approx 0.020\,\mpl$ below which inflation would fail. This corresponds to $r_{\text{crit}} \approx 5.6\times 10^{-6}$, but again remember that the precise value depends on the number of \efolds assumed for the homogeneous solution.

The paper is organised as follows. In Section~\ref{sect:theory}, we introduce the models of inflation we consider and discuss the space of initial conditions we explore. In Section~\ref{sect:Numerical Results}, we discuss the numerical results. We conclude in Section~\ref{sect: outlook}. We set $c=\hbar=1$ throughout, and work with the non-reduced Planck mass $\mpl=G^{-1/2}$.

\section{Theory and Methodology}\label{sect:theory}
\subsection{Model Space}

To proceed, we assume that at the beginning of inflation, the system is well described by a single scalar field minimally coupled to gravity
\begin{equation}
    S = \int \dd^4 x \sqrt{-g}\left(\frac{\mpl^2}{16\pi }R - \frac{1}{2}\partial_{\mu}\phi\partial^{\mu}\phi - V(\phi)\right)\,.
\end{equation}
We focus our study on an $\alpha$-attractor model~\cite{Kallosh:2013hoa,Kallosh:2014xwa} motivated from supergravity and string theory, and in good agreement with current observational data \cite{Planck:2018jri}.

\begin{figure}[htbp]
\centering
\includegraphics[width=1.04\columnwidth]{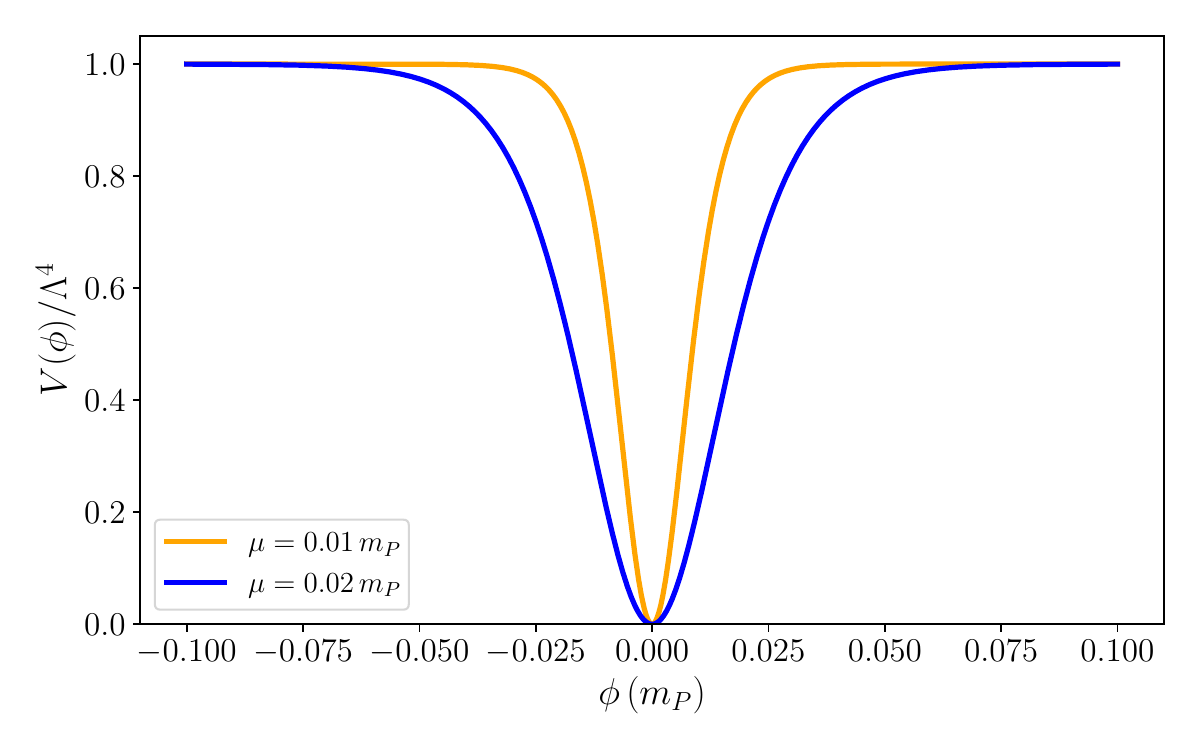}
\caption{The dependence of the shape of the potential $V(\phi)$ for $T$-model $\alpha$-attractor with respect to $\mu$ values.}
\label{fig:tanh_potential}
\end{figure}

The $\alpha$-attractor models possess an exponential plateau region where the scalar field can slow roll to a stable minimum at $\phi=0$.  They are split into two categories: the $E$-model and the $T$-model. For the $T$-model, its potential is 
\begin{equation}
    V(\phi) = \Lambda^4\tanh^2\left(\frac{\phi}{\mu}\right) \label{eqn:T_pot} \,,
\end{equation}
which is symmetric around $\phi=0$. Here $\Lambda$ sets the energy scale of inflation and is determined by the amplitude of the scalar power spectrum, and $\mu$ is its characteristic scale which describes how steep the exponential plateau at both sides of the well is. As can be seen in Figure~\ref{fig:tanh_potential}, the \emph{lower} the characteristic scale $\mu$, the steeper the potential, and hence the less robust inflation is to field inhomogeneities which can drag the inflaton down to the minimum.

For the $T$-model, the relation between the characteristic scale and $r$ is given by~\cite{Kallosh:2013yoa}
\begin{align} 
    r= \frac{16\pi\mu^2}{\mpl^2N_{*}^2}\,, \label{eqn:r}
\end{align}
where $N_*$ denotes the number of \efolds before the end of inflation at which the CMB pivot scale exits the horizon. In practice, $N_*$ depends on the details of reheating, and the choice of pivot scale, and is typically in the range $50-60$ \efolds. For example, for a pivot scale of $k_*=0.05$ Mpc$^{-1}$, instantaneous reheating gives $N_*\approx 57$, while a pivot scale of $k_*=0.002$ Mpc$^{-1}$ with instantaneous reheating gives $N_*\approx 60$. 
This means that we need around $N_*\approx 60$ \efolds of inflation for instantaneous reheating, and we will use this number to determine whether a given model with inhomogeneous initial conditions successfully inflated or not.

Meanwhile, the value $\Lambda^4$ is determined for a fixed parameter $\mu$ by the observed amplitude of the scalar power spectrum. 
\begin{equation}
\Delta_{\mathcal{R}}^2=\frac{H^2}{\pi \mpl^2 \epsilon_*} \,,
\end{equation}
where the slow-roll parameter
\begin{equation}
    \epsilon = \frac{\mpl^2}{16 \pi}\left(\frac{V^{\prime}}{V}\right)^2 ,
\end{equation}
is evaluated at the time when modes corresponding to the CMB pivot scale exit the horizon.

Thus for the $T$-model one obtains
\begin{equation}
    \Delta_{\mathcal{R}}^2(\phi_{*}) = \frac{32 \pi  \Lambda^4}{3\mpl^4} \frac{\mu^2}{\mpl^2} \sinh^4\left(\frac{\phi_{*}}{\mu}\right)\,,
\end{equation}
where $\phi_{*}$ is the field value at the time when modes whose wavenumber is given by the CMB pivot scale exit the horizon.

Let us also briefly comment on the $E$-model of the $\alpha$-attractor family, which has the potential 
\begin{equation}
    V(\phi) = \Lambda^4\left(1-e^{\phi/\mu}\right)^2\,,
\end{equation}  
for which $V\rightarrow \Lambda^4$ if $\phi\ll 0$ and $V\sim e^{2\phi/\mu}$ if $\phi>0$. Si= 0nce this potential increases exponentially past its minimum, at low $\mu$, large inhomogeneous initial conditions can and often lead to large potential energies as  in many regions of space the inflaton starts at this large potential regime. In the way we choose initial conditions, this leads to numerical issues caused by local exponentially large energy densities so that we did not investigate this model. 

\subsection{Initial Scalar Data}

As discussed in the introduction,  in the absence of a theory of initial conditions for inflation,  we model the scalar field with configurations where the volume-averaged initial gradient energy density and volume-averaged kinetic energy are of the same order as the potential energy i.e. $\langle\rho_{\nabla\,0}\rangle \approx \langle\rho_{K\,0}\rangle\approx   V(\phi_0)$, where $V(\phi_0)$ represents the energy scale associated with the homogeneous case at the scale $\phi_0$ where inflation starts.  Here the spatial volume average for any quantity $X$ on the hypersurface at fixed time $t$ is defined as
\begin{equation}
    \langle X\rangle \equiv\frac{\int_\Sigma d^3x\sqrt{\gamma}X}{\int_\Sigma d^3x\sqrt{\gamma}}\,.
\end{equation}

We consider harmonic \textit{pseudo-isotropic} perturbations of the scalar field in all three spatial directions, with the homogeneous background canonical momentum initially set to zero:
\begin{align}
\phi_\mathrm{init}(\textbf{x}) & = \phi_0 + \frac{\Delta \phi}{3} \sum_{i=1}^{3} \cos{\left(\frac{2\pi  x_i}{L}\right)}\,, \label{eqn:SF_profile} \\
\Pi_\mathrm{init}(\textbf{x}) & =\frac{\Delta \Pi}{3} \sum_{i=1}^{3} \cos{\left(\frac{2\pi x_i }{L}+ \theta\right)}\label{eqn:SF_mom_profile}\,.
\end{align}
and setting periodic boundary conditions. We choose $\phi_0$  such that, in the absence of scalar inhomogeneities and with the kinetic sector vanishing, inflation would last for 100 e-folds.  In models where the ``inflationary plateau'' extends to large field values, setting $\phi_0$ further from the reheating region naturally enhances the robustness of inflation. The parameters $\Delta \phi$ and $\Delta \Pi$ define the amplitude of the scalar field and its momentum inhomogeneities, while $\theta$ is the relative phase between the perturbations.  The length scale $L$ is the simulation box size, chosen to match the initial Hubble radius $H_0^{-1}$ in the absence of inhomogeneities:
\begin{equation}
L = \frac{3 \mpl}{\sqrt{24 \pi V\left(\phi_0\right)}} \label{eqn:hubble_radius}\,.
\end{equation}

For our initial data \eqn{eqn:SF_profile} and \eqn{eqn:SF_mom_profile}, the volume averages are
\begin{align}
\langle\rho_{\nabla, 0}\rangle &= \langle\frac{1}{2}(\nabla \phi_{\mathrm{init}})^2\rangle = \frac{\pi^2  \Delta\phi^2}{3L^2} \,,
\\
\langle\rho_{K, 0}\rangle &= \langle\frac{1}{2}\Pi_{\mathrm{init}}^2\rangle = \frac{ \Delta\Pi^2}{12} \,. 
\end{align}
We set the values of $\Delta \phi$ and $\Delta \Pi$ such that $\langle\rho_{\nabla, 0}\rangle=\langle\rho_{K, 0}\rangle=V(\phi_0)$. We will make several remarks regarding our choices beyond that of its densities as follows.

First, we have chosen to pick a single mode of wavelength $L=H_0^{-1}$. The reason is previous studies \cite{Clough:2016ymm,Clough:2017efm,Aurrekoetxea:2019fhr} have consistently shown that modes that are close to horizon-size are the most dangerous for inflation, since equal energy sub-horizon modes generally result in a smaller displacement of the scalar field from its mean $\phi_0$ and decay faster, while super-horizon modes simply change the background values. 

Second, as mentioned before, we choose the mean value $\phi_0$ such that it will generate 100 $e$-folds. For the $\alpha$-attractor $T$-model, since the plateau region extends to infinity there is no \emph{a priori} reason that $\phi_0$ should be anywhere near this value. Moreover, depending on the physics that preceded inflation  $\phi_0$ could be larger, which would lower the critical value of $\mu$ and $r$. 

Third, we chose not to add background initial momentum. In general, if the inflationary potential is not symmetric (e.g., an $E$-model or a Coleman-Weinberg-like potential), then adding initial homogeneous momentum generally makes inflation less robust \cite{Corman:2022alv,Elley_2025}. However, since in our case the $T$-model is symmetric, homogeneous initial momentum can push the inflaton across the minimum over to the other side of (still inflationary) potential, which will then roll down and inflate. This has the surprising effect of making inflation more robust for large homogeneous initial momentum. We have chosen to set it to zero to simplify our analysis.

\subsection{Initial Geometric Data}

In order to perform numerical simulations, we  foliate spacetime into spatial hypersurfaces that evolve along a time direction via the  standard ADM metric:
\begin{equation}
\dd s^2 = -\alpha^2\dd t^2 + \gamma_{ij}(\dd x^i + \beta^i \dd t)(\dd x^j + \beta^j \dd t)\,,
\end{equation}
where $\gamma_{ij}$ is the 3-dimensional spatial metric of the hypersurfaces, which we evolve along with the extrinsic curvature $K_{ij} = \partial_t \gamma_{i j} + 2 D_{(i} \beta_{j)}$. The spatial metric and the extrinsic curvature are the physical geometric degrees of freedom. The lapse function $\alpha$ and shift vector $\beta^i$ are evolved as gauge choices, reflecting the coordinate freedom in GR. We further decompose the extrinsic curvature as
\begin{equation}
K_{ij} = A_{ij} + \frac{1}{3}\gamma_{ij}K\,,
\end{equation}
for which  $K = \gamma^{ij} K_{ij}$ is the trace and $A_{ij}$ is the traceless part. The trace $K$ is associated with the divergence or convergence of a congruence of geodesics and thus measures local expansion or collapse, with our convention being that $K < 0$ corresponds to expansion, while $K > 0$ indicates collapse. The traceless part $A_{ij}$ is related to tensor modes and, in the low-energy limit, can be interpreted as a gravitational wave background. We further extract the conformal factor $\chi$ from the spatial metric to obtain the conformally related metric $\gamma_{ij} = \chi^{-1} \tilde{\gamma}_{ij}$.

Initial data must satisfy the Hamiltonian and momentum constraints of General Relativity. Thus, given an initial matter configuration and some initial geometric data for the hypersurface, we solve these constraints to determine the remaining geometric degrees of freedom. We do this using the CTTK method \cite{Aurrekoetxea:2022mpw,Aurrekoetxea:2025kmm}, where we assume an initially conformally flat metric $\tilde{\gamma}_{ij} = \delta_{ij}$ and fix the conformal factor to $\chi = 1$. This leads to an algebraic equation for the Hamiltonian constraint and a Poisson-like equation for the momentum constraint, which are solved to obtain the initial profiles of the expansion $K$ and trace-free extrinsic curvature $A_{ij}$, respectively; see \cite{Aurrekoetxea:2022mpw} for details.

Note that this slicing choice implies that the intrinsic spatial curvature $^{(3)}R = 0$, which is a special case; it has been suggested that other choices may yield significantly different outcomes \cite{Garfinkle:2023vzf}, but this is currently contested (see \cite{Joana:2024ltg}). 

We use periodic boundary conditions, which enforce integrability conditions on the constraint equations. Specifically, the integral of the constraints over the domain must vanish. With an initially constant conformal factor, this yields the condition $\int S_i \dd V = 0$, $S_i$ being the momentum density. This condition  is satisfied as long as the perturbation modes are also periodic. We have shown in \cite{Elley_2025} that the phase $\theta$ between the spatial and kinetic perturbations can impact the robustness of inflation depending on whether the kinetic perturbations reinforce/suppress the motion of the inflaton towards the reheating minimum. However, in the limit of large kinetic energy densities, we expect that for both in-phase and out-phase kinetic inhomogeneities, the maximum number of \efolds{} should be reduced. In this work, we check for both in-phase $\theta = 0$ and out-of-phase $\theta = \pi$ perturbations.

\subsection{Evolution}

To evolve the spacetime quantities we use the CCZ4 formulation \cite{Dumbser_2018}, which provides constraint damping and therefore stability for simulations with large gradients.  For details, see Appendix \ref{appendix:ccz4}.

To drive the lapse $\alpha$, we use the ``cosmologically averaged'' driver
first introduced by \cite{Giblin:2019nuv}, with the inclusion of the relevant CCZ4 term
\begin{equation}
\partial_t \alpha = - \alpha \left( K - \langle K \rangle -2\Theta \right) + \beta^i \partial_i \alpha\,,
\end{equation}
while the shift vector $\beta^i$ is driven using the standard puncture gauge \cite{Campanelli:2005dd,Baker:2005vv}. Here $\langle K \rangle$ is the volume average of the (trace) extrinsic curvature. 
Including the volume average $\langle K \rangle$ in the lapse evolution prevents the lapse from growing excessively in expanding regions ($K < 0$), while recovering the standard moving puncture gauge in collapsing regions. This mitigates the formation of steep gradients and discontinuities in $K$, which would otherwise lead to large constraint violations due to regions where the lapse function becomes significantly larger than in neighboring regions. 

\subsection{Diagnostics}

\begin{figure*}[t!]
        \includegraphics[width=1.0\textwidth]{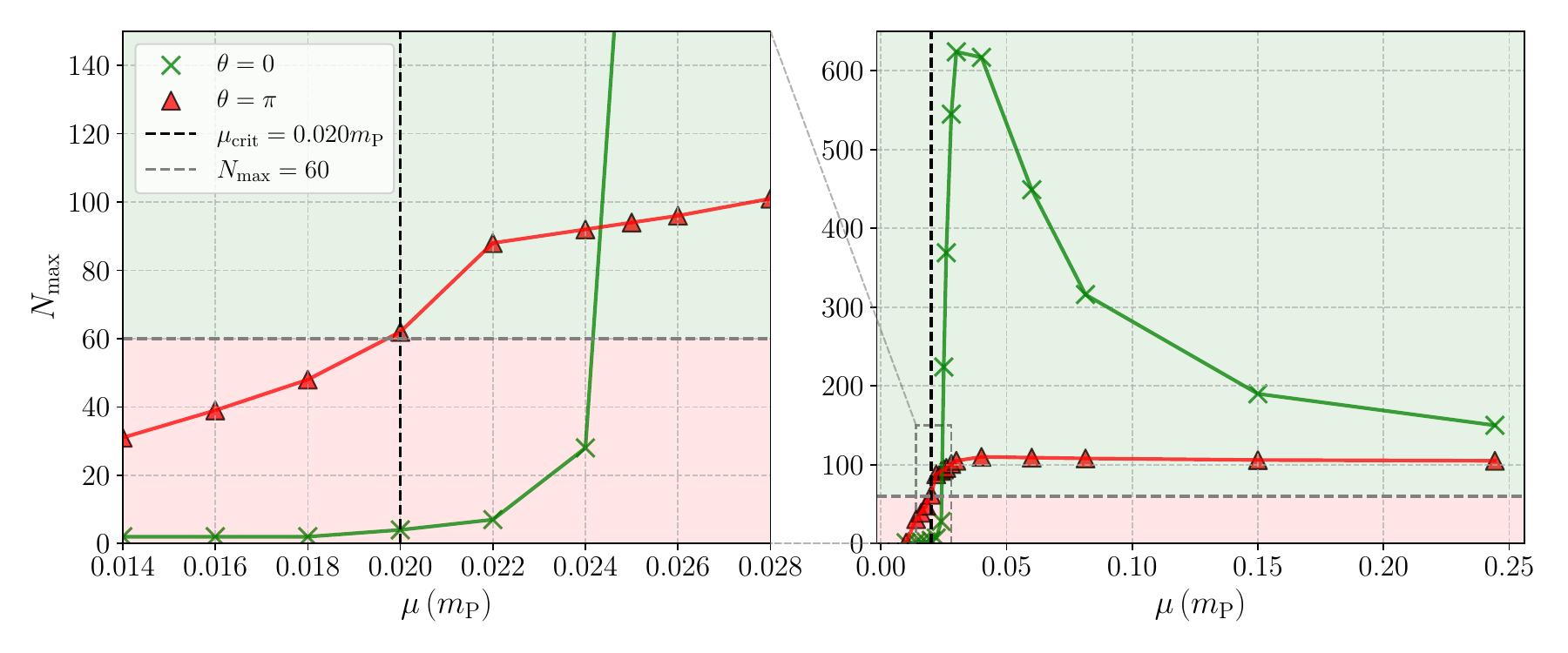}
    \caption{Dependence of the maximum number of $e$-folds, $N_{\text{max}}$, on the characteristic scale $\mu$, with initial conditions $\langle\rho_{\nabla,0}\rangle = \langle\rho_{K,0}\rangle = V(\phi_0)$. The red-shaded area indicate a maximum number of \efolds{} $N_{\text{max}}<60$ which we define as failure of inflation, whereas the green-shaded area shows $N_{\text{max}}>60$ and represents successful inflation. Here the red branch corresponds to out-of-phase initial perturbations ($\theta = \pi$), and the green branch to in-phase ones ($\theta = 0$). The left panel shows the results for $0.014\mpl\leq\mu < 0.026\mpl$ and the right panel for $0.014\mpl\leq\mu \leq \sqrt{3/16\pi}\mpl$. We observe that the critical values for which there is a successful amount of inflation are $\mu = 0.02\mpl$ and $\mu = 0.025\mpl$ for the out-of-phase and in-phase perturbations respectively. These correspond to the critical tensor-to-scalar ratios $r = 5.6\times10^{-6}$  and $r = 8.7\times10^{-6}$.}
    \label{fig:nmax_mu_1}
\end{figure*}

In the FLRW limit, the Hamiltonian constraint reduces to the Friedmann equation, and the trace of the extrinsic curvature is related to the Hubble parameter $K=-3H$ while the scale factor by the conformal factor $a(t) = \chi^{-1/2}$. Thus, the number of \efolds{} is given by
\begin{equation}
    N = -\frac{1}{2}\ln\chi\,.
\end{equation}
Most of the dynamics occurs within the first few $e$-folds, after which the inhomogeneities exit the horizon (if inflation gets started), and subsequently slowly rolls down the potential. A typical simulation leads to disjoint regions of space, with some regions where the scalar field has reached the reheating minimum (ending inflation) and some regions in which the scalar field is approximately homogeneous and on the inflating plateau.  After some period of evolution, and if some region of $H_0^{-1}$ homogenizes sufficiently, we extract $\phi$ and $\dot{\phi}$ at this region and extrapolate the expected maximum number of \efolds{} by using the FLRW equations of motion
\begin{align}
    \ddot{\phi} &+3H\dot{\phi} + V'(\phi) = 0\,, \label{eq:hom_KG} \\
    \left( \frac{\dot{a}}{a} \right)^2 & = \frac{8\pi }{3\mpl^2}\left(\frac{1}{2}\dot{\phi}^2 + V(\phi)\right)\,.\label{eq:hom_FRW= 0}
\end{align}
We can use the homogeneous equations as these regions have spatial dimensions equal to or exceeding the Hubble radius associated with the inflationary vacuum  $\geq H_{0}^{-1}$. We declare that inflation has succeeded if there is such a homogenised patch that gives more than $60$ \efolds{} of inflation as the field rolls from the extracted value until the value of $\phi = \phi_{\text{end}}$ where $\epsilon = 1$.
For the $T$-model, this is
\begin{equation}
    \phi_{\text{end}} = \frac{\mu}{2}\sinh^{-1}\left(\pm\frac{\mpl}{\mu\sqrt{\pi}}\right)\,,
\end{equation}
where the $\pm$ accounts for the fact that there are two inflationary plateaus, and for the maximum number of \efolds{}\footnote{Another diagnostic used in these studies is the volume averaged number of \efolds{} which is defined as $\langle N\rangle = -1/2\langle\ln\chi\rangle$. This value is  dominated by the regions which continue inflating.} we choose the one where the slow-roll occurs. The scalar-tensor ratio $r$ is then calculated using \eqn{eqn:r}.

\section{Numerical Results}\label{sect:Numerical Results}

\subsection{Results for $\langle\rho_{K\,0}\rangle = \langle\rho_{\nabla\,0}\rangle = V(\phi_0)$}

\begin{figure*}[t]{
\centering
\includegraphics[width=\textwidth]{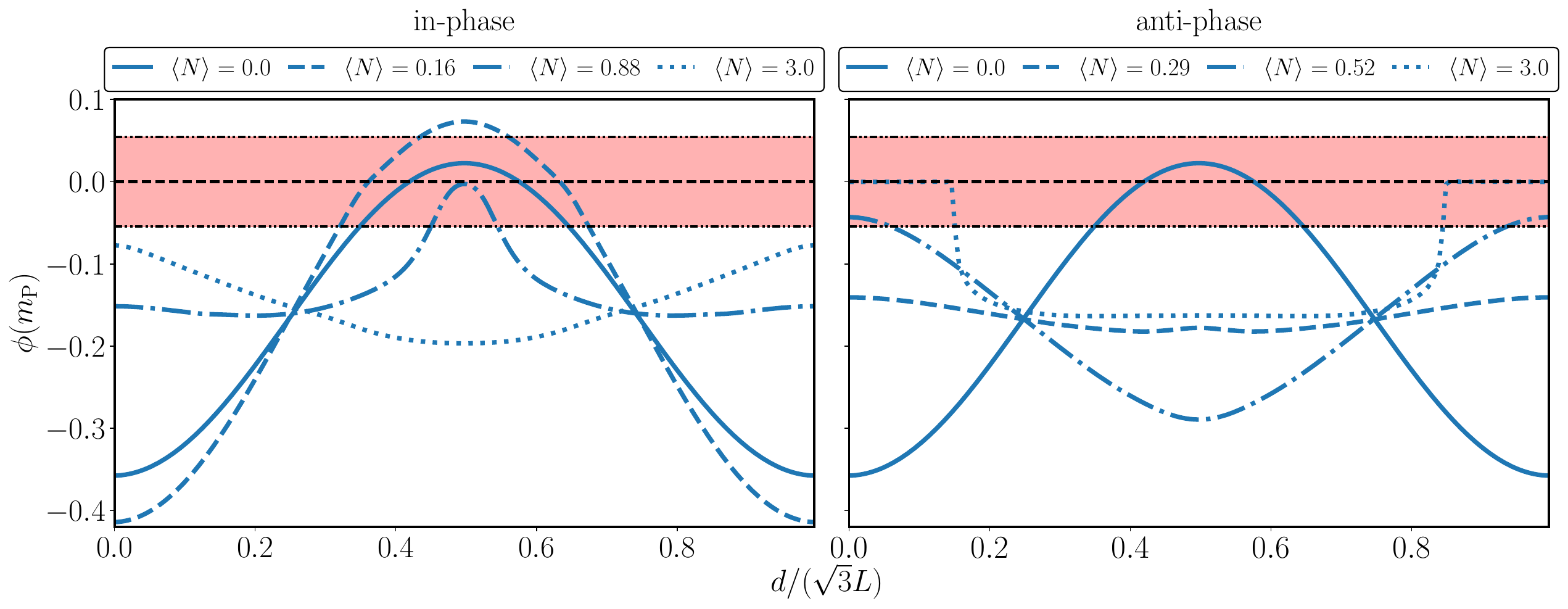}
}
\caption{Evolution of the field profiles across the largest diagonal of the simulation $(0,0,0)\xrightarrow{}(L,L,L)$ for both the in-phase (left) and out-of-phase (right) initial field configurations with $\mu = 0.03 \mpl$. The shaded red region indicates the field values for which the potential cannot support slow-roll i.e. $\epsilon \geq 1$. For both cases we observe inflation succeeding in the centre but failing at the corners. However, for the in-phase case the central field initially explores higher up the positive-$\phi$ side of the potential, before being slingshot up the plateau by steep restorative field gradients to subsequently yield $> 600$ \efolds of inflation.   }
\label{fig:phi_profile_evol}
\end{figure*}

We vary the  characteristic scale $\mu$ between $0.01\mpl \leq \mu \leq \sqrt{3/16\pi}\mpl$ corresponding\footnote{Notice that the maximum examined value $\mu = \sqrt{3/16\pi}\mpl$ corresponds to the Starobinsky scale of the $E$-model. However, for the $T$-model this value is not a special scale. We instead consider it as an upper limit, such that models with near-Planckian characteristic scales are included.} to $1.4\times10^{-6}\leq r\leq8.3\times10^{-4}$. We fix both the gradient and kinetic inhomogeneities such that $\langle\rho_{K,0}\rangle = \langle\rho_{\nabla,0}\rangle = V(\phi_0)$, motivated by the equipartition argument discussed in the previous section. The kinetic inhomogeneities are specified to be either in-phase with the corresponding gradient ones (phase $\theta = 0$) or out-of-phase (phase $\theta = \pi$).

Our results are presented in Figure~\ref{fig:nmax_mu_1}. We observe an increase in $N_{\rm max}$ with $\mu$ for both in-phase and out-of-phase perturbations. The transition between unsuccessful and successful inflation - defined by  $N_{\text{max}} \approx 60$ - occurs at $\mu_{\mathrm{crit}} \approx 0.025\,\mpl$ and $\mu_{\mathrm{crit}} \approx 0.020\,\mpl$ for the in-phase and out-of-phase initial perturbations respectively. Thus, using \eqn{eqn:r} the corresponding critical tensor-to-scalar ratios are $r_{\text{crit}} \approx 5.6\times10^{-6}$  and $r_{\text{crit}} \approx 8.7\times10^{-6}$.

For small values of $\mu$ i.e. $\mu < 0.02 \mpl$ we do not obtain the requisite $N_{\mathrm{max}} = 60$ for any of the simulations and therefore deem these failures. The failure occurs as the perturbations push the field towards the minimum, which causes some regions to fall in and subsequently drag the rest of the field down. Increasing $\mu$ both shifts the field higher up the plateau and also reduces the potential gradient near the minimum, thus reducing the loss of \efolds{}. This is a well-known result from many numerical simulations of inflationary spacetimes \cite{East:2015ggf,Clough:2016ymm,Aurrekoetxea:2019fhr,Corman:2022alv,Joana:2020rxm,Joana:2020rxm,Elley_2025} -- higher characteristic scale inflation is generally more robust to inhomogeneities.

As discussed in our previous paper \cite{Elley_2025},  in-phase perturbations result in a lower number of \efolds{} than their out-of-phase counterparts for small $\mu$ values, even preventing inflation entirely for $\mu \leq 0.018 \mpl$. One can explain this by noting that for the out-of-phase cases, the direction of momentum acts to homogenise the field. On the other hand, for in-phase perturbations the field extrema are initially moving away from each other in field space, resulting in a rapid growth in the field gradients. As the regions of the field close to the minimum get stuck due to high potential gradients, the field that is furthest up the plateau is slingshot back towards the minimum by gradient pressure, drastically reducing $N_{\mathrm{max}}$. 
 
For large $\mu$ we see $N_{\text{max}}$ begin to plateau at around $N_{\text{max}} = 100$ for both initial configurations, which is what we would have expected without the inhomogeneities. This is because  the field is much further up the plateau, and the slope of the potential is shallower, so the potential gradient no longer plays a significant role in the dynamics. The perturbations simply decay due to the restoring force from the field gradients and inflate normally.

\begin{figure*}[t!]{
\includegraphics[width=0.7\textwidth]{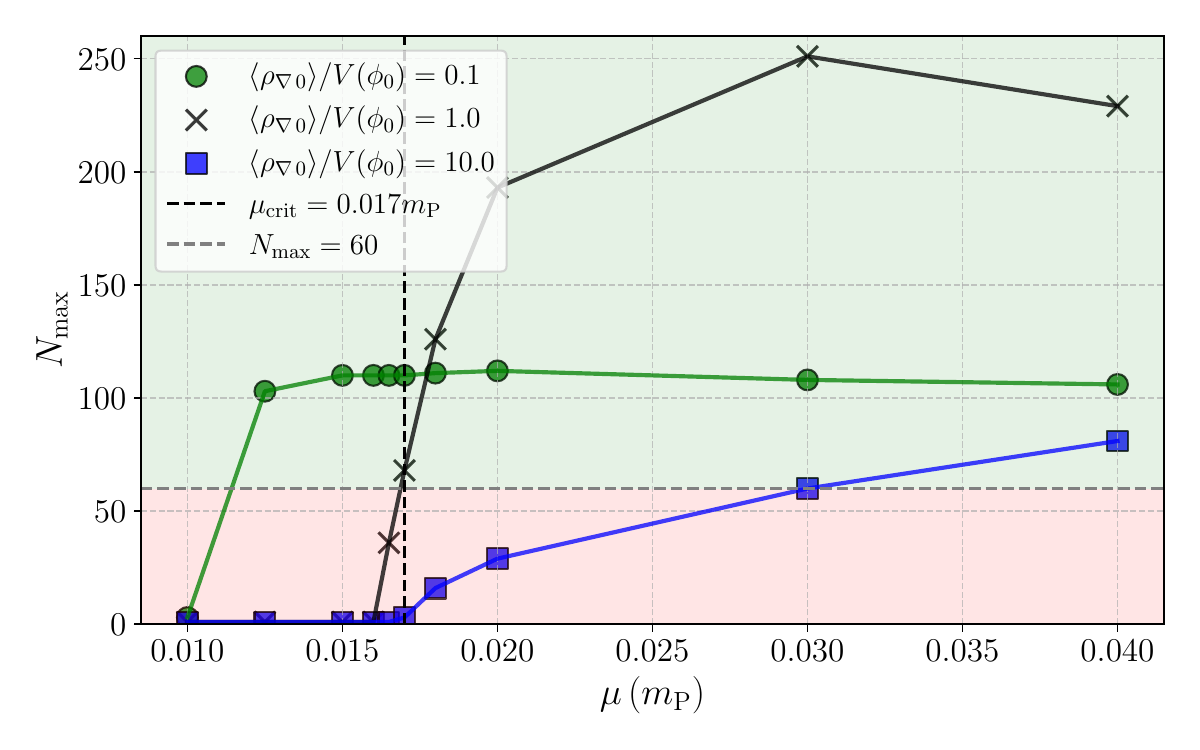}
}
\caption{Results for the dependence of the maximum number of \efolds{} $N_{\text{max}}$ on the characteristic scale $\mu$ for the case without kinetic inhomogeneities. The red-shaded area shows a maximum number of \efolds{} $N_{\text{max}}<60$ which we define as failure, whereas the green-shaded area shows $N_{\text{max}}>60$ and represents successful inflation. Each dot  shows the $N_{\text{max}}$ of that simulation, and we present with different colours three branches of fixed ratio $\langle \rho_{\nabla\,0}\rangle/V(\phi_0)$. For $\mu \gtrsim 0.017\mpl$ simulations with initial conditions such that $\langle\rho_{\nabla}\rangle= V(\phi_0)$ (black line) give $N_{\text{max}}>60$. This suggests a smaller lower bound on $\mu$ and thus to the $r_{\text{crit}} = 4.0\times10^{-6}$ compared to the case where also kinetic inhomogeneities are included. The blue branch represents initial conditions such that $\langle\rho_{\nabla\,0}\rangle =10V(\phi_0)$ and we obtain successul inflation for $\mu\geq0.03\mpl$ which gives $r_{\text{crit}} = 1.3\times10^{-5}$. We note that this seemingly monotonic behaviour between energy ratios and $\mu_{\text{crit}}$ holds only over large energy ranges -- small differences in energy ratios around unity do not necessarily lead to monotonic behaviour due to the presence of the minimum in the potential. }
\label{fig: nmax_mu}
\end{figure*}

In between these small characteristic scales, for $0.024\mpl < \mu \leq 0.03\mpl$, we see an interesting phenomenon where there is a sudden spike in $N_{\mathrm{max}}$ for the in-phase case, reaching values up to $N_{\mathrm{max}} = 619$ -- \emph{initial perturbations make inflation even more robust than the initially homogeneous case!} This can be understood from the evolution of the profile of the field across the largest diagonal of our grid is shown in the left panel of Figure~\ref{fig:phi_profile_evol}. We see that in the central region, for which initially $\phi = \phi_{\mathrm{max}}$, the field is in the reheating minimum. As a result of the initial field velocity acting to exacerbate the field gradients, during the evolution the gradients near this region become so large that they pull the field from the minimum and slingshot it up onto the plateau, further than $\phi_0$. This only becomes possible once the potential gradient is sufficiently shallow and enough of the field is initially higher up the plateau. So for this range of $\mu$, initial $\phi_{\mathrm{min}}$/$\phi_{\mathrm{max}}$ end up swapping places, as each is propelled by the field pressure towards/away from the potential minimum respectively.  This phenomenon is not present in the out-of-phase case as the initial momentum pushes the field in the opposite direction, helping homogenise the field as shown in the right panel of Figure~\ref{fig:phi_profile_evol} (though it tends to overshoot).   

This prompted us to consider background values $\phi_0$ such that in the homogeneous limit inflation would have failed (i.e. yield $N_{\text{max}} < 60$), and yet by introducing large inhomogeneities some patches can enter an attractor phase and produce prolonged inflation. We tested such an example for $\mu = 0.03\mpl$ with $\phi_0$ fixed so that, in the absence of inhomogeneities, it yields $30$ \efolds{}. We find that the central patch of the simulation domain is indeed pulled back to the left plateau as described above, and produces $N_{\text{max}} = 81$ \efolds{}indicating successful inflation. The conclusion is that certain configurations of large inhomogeneities can in fact lead to longer inflation. Setting such $\phi_0$ with scalar and kinetic inhomogeneities should give alternative constraints on $r$ which are necessary in a fully comprehensive study of the initial condition phase space for the $T$-model. We leave such configurations for future work.

\subsection{Results for $\langle\rho_{\nabla\,0}\rangle = V(\phi_0)$ and $\langle\rho_{K\,0}\rangle = 0$}

For completeness we studied the dependence of $N_{\mathrm{max}}$ on field inhomogeneities only (i.e. $\Delta\Pi = 0$).  The results of these simulations are summarized in Figure~\ref{fig: nmax_mu}.

First, we ran simulations for different values of $\mu$ in the limit of small characteristic scale, where the distance $\delta\phi$ remains sub-Planckian. The $\mu$ values we choose to work with are in the range $0.01 \mpl\leq\mu\leq 0.04 \mpl$ which correspond through \eqn{eqn:r} to tensor to scalar ratios $1.4\times 10^{-6}\leq r_{\text{crit}}\leq2.2\times10^{-5}$. Here $\mu_{\text{crit}}$ here is smaller, which corresponds to a lower value of $r_{\text{crit}}$. This is expected given the absence of initial momentum fluctuations.

We also explore how $\mu_{\text{crit}}$ changes for different values of $\langle \rho_{\nabla,0}\rangle/V(\phi_0) = 0.1,1,10$. As can be seen in Figure~\ref{fig: nmax_mu}, the general result is that larger gradient energy densities lead to an increase in $r_{\text{crit}}$. We hasten to add that this seemingly monotonic relationship between $\langle \rho_{\nabla,0}\rangle/V(\phi_0)$ and $r_{\text{crit}}$ (i.e. the value of $r$ when each line crosses $N_{\mathrm{max}} = 60$) does not strictly hold in the regime when $\langle \rho_{\nabla,0} \rangle /V(\phi_0)\sim 1$, since the presence of the minimum in the potential leads to local effects that may dominate the dynamics (i.e. the slingshot mechanism discussed earlier).

From these results we obtain $\mu_{\text{crit}} = 0.017\mpl$, where $N_{\text{max}}>60$ in Figure~\ref{fig: nmax_mu} for values $\langle\rho_{\nabla\,0}\rangle = V(\phi_0)$. Consequently,  the relevant lower bound is $r_{\text{crit}}= 4.0\times 10^{-6}$. This result is consistent with the result of the previous section - since kinetic perturbations make inflation less robust for small values of $\mu$, we expected in their absence the critical value of $\mu$ to be lower compared to the value $\mu = 0.020\mpl$ we obtained. However, the overall conclusion remains and for initial conditions for which the homogeneous solution achieves $100$ $e$-folds, the tensor-to-scalar ratio remains $r_{\text{crit}}\sim 10^{-6}$.  

\section{Summary and Outlook}\label{sect: outlook}

We have shown that for an $\alpha$-attractor $T$-model, with initial conditions such that the volume-averaged gradient, kinetic and potential energies are approximately equal, and for a given number of \efolds for the homogeneous solution, there exists a critical characteristic scale below which inflation fails. Assuming $100$ \efolds for the homogeneous solution, inflation fails to inflate for $60$ \efolds below $\mu_{\text{crit}}\lesssim 0.02\mpl$. Since the scalar-to-tensor ratio of the model is directly dependent on $\mu$, this translates to a $r_{\text{crit}}\gtrsim 5.6\times 10^{-6}$. As stronger inhomogeneous initial conditions require a higher characteristic scale for sufficient inflation, one generically expects that (a) increasing the amplitudes of the initial perturbations increases $r_{\text{crit}}$, while (b) removing kinetic perturbations decreases $r_{\text{crit}}$. We confirmed these expectations with numerical simulations.

Let us make two comments regarding our results. First, for a given energy density of the perturbations, we only simulated horizon-scale modes, as these are the most effective in ending inflation \cite{Clough:2016ymm,Joana:2020rxm,Aurrekoetxea:2019fhr,Elley_2025}. Nevertheless, a true interpretation of ``equipartition'' would mean that this and all subhorizon modes contribute to the total energy density. Such a configuration is unfortunately extremely expensive to simulate numerically. Given sub-horizon modes tend to be less effective at ending inflation, by only including horizon-sized modes we erred on the side of making the initial conditions more dangerous for inflation than they might have been, meaning that for true equipartition, $\mu_{\text{crit}}$ is likely \emph{smaller} than what we have found.

Second, we have arbitrarily set the mean value $\phi_0$ of the initial perturbations to be such that inflation would have yielded 100 \efolds in the homogeneous limit. For a $T$-model, the inflationary plateau naively extends to infinity, so that $\phi_0$ could be taken sufficiently large such that any initial perturbations would have inflated away~\cite{East:2015ggf, Clough:2016ymm,Aurrekoetxea:2019fhr}. Thus, there is only a critical value for $\mu$ and $r$ if $\phi_0$ is held fixed. That said, there are good reasons to believe that the system is not described by a single scalar field with a flat potential over an infinite range in field space~\cite{Obied:2018sgi} so that there is presumably some lower bound. It is interesting to ask whether such bounds exist and to what extent it could provide targets for future observations. 

In this paper, we demonstrate a proof of principle of exactly how such a bound might arise from the interplay between the inflationary model space and the phase space of possible initial conditions. It is natural to ask whether there are similar bounds on other cosmological observables.
   
\section{Acknowledgements}
We would like to thank Cheng Cheng, Nicole Righi, Josu Aurrekoetxea, Katy Clough,  Santiago Agüí Salcedo, Daniel G. Figueroa and Maximilian Detering for their useful input. We would also like to thank members of the GRTL Collaboration \href{http://www.grtlcollaboration.org}{(http://www.grtlcollaboration.org/)}. This work is supported by a Research Project Grant RPG-2021-423 from Leverhulme Trust.  ME has been supported in part by the PID2021-123703NB-C21 grant funded by MCIN/AEI/10.13039/501100011033/and by ERDF; “A way of making Europe”; the Basque Government grant (IT-1628-22). RF is also supported in part by a US Department of Energy Grant, DE-SC0009919. This work used the DiRAC Memory Intensive service (Cosma8 / Cosma7 / Cosma6 [*]) at Durham University, managed by the Institute for Computational Cosmology on behalf of the STFC DiRAC HPC Facility (www.dirac.ac.uk). The DiRAC service at Durham was funded by BEIS, UKRI and STFC capital funding, Durham University and STFC operations grants. DiRAC is part of the UKRI Digital Research Infrastructure.

\bibliographystyle{apsrev4-1}
\bibliography{mybib.bib}
\appendix

\section{Evolution Equations}\label{appendix:ccz4}
For the scalar-field evolution, the 3+1 decomposition of the scalar-field equation of motion $\nabla_{\mu}\phi\nabla^{\mu}\phi-V''(\phi) = 0$ yields
\begin{align}
\partial_t \phi &= \alpha \Pi +\beta^i\partial_i \phi \label{eqn:dtphi2} \,, \\
\partial_t \Pi&=\beta^i\partial_i \Pi + \alpha\partial_i\partial^i \phi + \partial_i \phi\partial^i \alpha \notag \\  &\hspace{0.4cm} +\alpha\left(K\Pi-\gamma^{ij}\Gamma^k_{ij}\partial_k \phi-\frac{dV}{d\phi}\right) \,,
\end{align}
with the canonical momentum of the scalar field $\phi$ to be
\begin{equation}
\Pi\equiv\frac{1}{\alpha}(\partial_t\phi-\beta^i\partial_i\phi)\,.
\end{equation}

The Hamiltonian and Momentum constraint equations are respectively:
\begin{align}
     \mathcal{H} &=  R^2+K^2 - K_{ij}K^{ij} -16\pi G\rho = 0\,, \\
    \mathcal{M}_i &=  D^j(\gamma_{ij} K - K_{ij})-8\pi G S_i = 0
\end{align}   
The geometric quantities are evolved according to the equations of the CCZ4 scheme:
\begin{equation}
   \partial_t\chi=\frac{2}{3}\chi\alpha K-\frac{2}{3}\chi\partial_k\beta^k+\beta^k\partial_k\chi\,,
\end{equation}
\begin{equation}
    \partial_t\tilde\gamma_{ij}=-2\alpha\tilde{A}_{ij}+2\tilde\gamma_{k(i}\partial_{j)}\beta^k-\frac{2}{3}\tilde\gamma_{ij}\partial_k\beta^k+\beta^k\partial_k\tilde\gamma_{ij}\,,
\end{equation}
\begin{equation}
\begin{split}
       \partial_tK&=-\gamma^{ij}D_iD_j\alpha+\alpha(R+2D_iZ^i+K^2-2\Theta K)\\&+\beta^j\partial_jK-3\alpha\kappa_1(1+\kappa_2)\Theta\,,
\end{split}
\end{equation}
\begin{equation}
\begin{split}
   \partial_t\tilde{A}_{ij}&=\chi[-D_iD_j\alpha+\alpha(R_{ij}-8\pi\alpha S_{ij})]^{TF}\\&+\alpha(K\tilde{A}_{ij}-2\tilde{A}_{il}{\tilde{A}^l}_j)+\tilde{A}_{ik}\partial_j
   \beta^k+\tilde{A}_{jk}\partial i\beta^k\\&-\frac{2}{3}\tilde{A}_{ij}\partial_k\beta^k+\beta^k\partial_k\tilde{A}_{ij}\,,
\end{split}
\end{equation}
\begin{equation}
\begin{split}
   \partial_t\Theta&=\frac{1}{2}\alpha(R+2D_iZ^i-\tilde{A}_{ij}\tilde{A}^{ij}+\frac{2}{3}K^2-2\Theta K)\\&-Z^i\partial_i\alpha+\beta^k\partial_k\Theta-\alpha\kappa_1(2+\kappa_2)\Theta\,,
\end{split}
\end{equation}
\begin{equation}
\begin{split}
    \partial_t\hat{\Gamma}^i&=2\alpha(\tilde{\Gamma}^i_{jk}\tilde{A}^{jk}-\frac{3}{2}\tilde{A}^{ij}\frac{\partial_j\chi}{\chi}-\frac{2}{3}\tilde\gamma^{ij}\partial_jK)-2\alpha\kappa_1\tilde\gamma^{ij}Z_j\\&+2\tilde{\gamma}^{ki}(\alpha\partial_k\Theta-\Theta\partial_ka-\frac{2}{3}\alpha KZ_k)-2\tilde{A}^{ij}\partial_j\alpha\\&+\beta^k\partial_k\hat{\Gamma}^i+\tilde{\gamma}^{kl}\partial_k\partial_l\beta^i+\frac{1}{3}\tilde{\gamma}^{iuk}\partial_k\partial_l\beta^l+\frac{2}{3}\tilde{\Gamma}^i\partial_k\beta^k\\&-\tilde\Gamma^k\partial_k\beta^i+2\kappa_3\left(\frac{2}{3}\tilde\gamma^{ij}Z_j\partial_k\beta^k-\tilde\gamma^{jk}Z_j\partial_k\beta^i\right)\,,
\end{split}
\end{equation}
where
\begin{equation}
   \hat{\Gamma}^i=-\partial_j\tilde{\gamma}^{ij}+2\frac{{\gamma^i}_\alpha Z^\alpha}{\chi}
\end{equation}
and
\begin{equation}
   \Theta\equiv Z^0\,.
\end{equation}
\section{Convergence Testing}

\begin{figure}[t!]
\centering
{\includegraphics[width=0.53\textwidth]{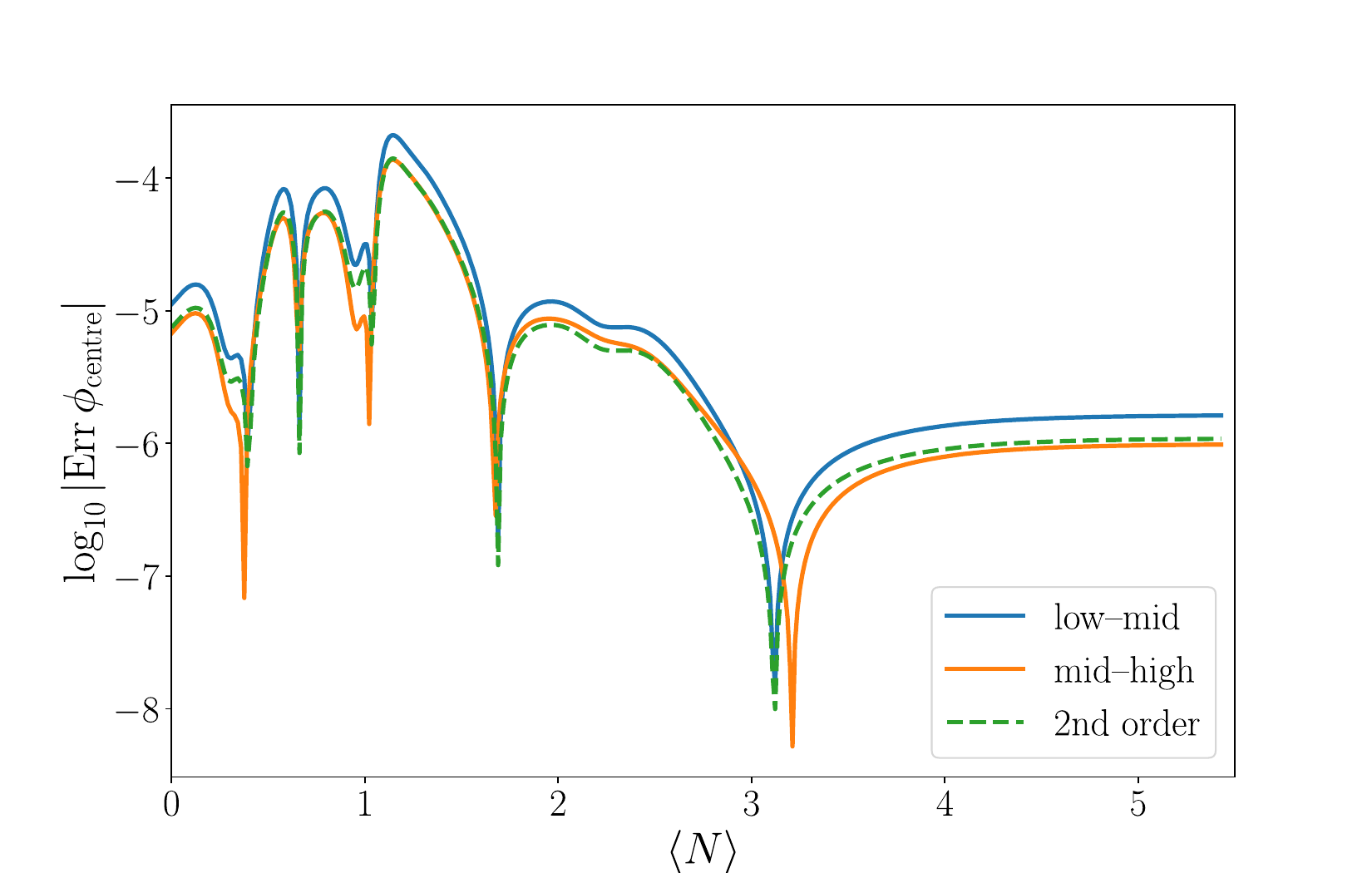}}
\caption{The covergence of the value of $\phi$ at the centre of the simulation domain for $\mu = 0.03\mpl$ and $\langle\rho_{\nabla\, 0}\rangle = \langle\rho_{K\,0}\rangle = V(\phi_0)$ and $\theta = 0$.} 
\label{fig:convergence}
\end{figure}

We consider our simulation domain to be a cubic box of size $L$ which is defined by~\eqn{eqn:hubble_radius}, with periodic boundary conditions and scalar inhomogeneity amplitudes $\Delta\phi/\mpl = \{0.06, 0.19, 0.6\}$ corresponding to ${\langle\rho_{\nabla}\rangle/V(\phi_0)} = \{0.1,1.0,10.0\}$.

We test the robustness of our numerical simulation by comparing the evolution for in three different base resolutions, namely $N_{\text{low}}=160$, $N_{\text{mid}}=192$, $N_{\text{high}}=224$,  for $\mu = 0.03\mpl$ and $\langle\rho_{\nabla\, 0}\rangle = \langle\rho_{K\,0}\rangle = V(\phi_0)$ and $\theta = 0$. 
In Figure~\ref{fig:convergence}, we show the convergence of the value of the scalar field at the centre of the simulation domain, which is used to extract the remaining number of \efolds{} whence the field has homogenised. This is defined as the absolute value of the differences between the base resolutions, and our analysis indicates second-order convergence. That happens at around $\langle N\rangle\approx 3$. 
\begin{equation}\label{converge}
    \frac{|\phi_{\text{centre}}^{mid}-\phi_{\text{centre}}^{\text{low}}|}{|\phi_{\text{centre}}^{\text{high}}-\phi_{\text{centre}}^{\text{mid}}|}\approx\frac{1/N_{\text{mid}}^p-1/N_{\text{low}}^p}{1/N_{\text{high}}^p-1/N_{\text{mid}}^p}
\end{equation}
where $p$ is the order of convergence~\cite{article}.

The volume average of the Hamiltonian constraint is well behaved until some regions stop inflating. In those regions, the scalar field oscillates around its potential minimum and these oscillations occur much faster compared to the Hubble time of inflation and we lose control over the constraints. However, we argue that those constraint violations are not spoiling our conclusions as the point we pick to extrapolate the remaining number of efolds has already been in attractor phase and it is causally disconnected with those regions that stop inflating and the local Hamiltonian constraint remains under control.

\section{Summary of model parameters}
\begin{table}[h!]
\centering
\begin{tabular}{|c|c|c|c|c|}
\hline
$\mu/\mpl$ & $\Lambda^4/\mpl^4$ & $\phi_0/\mpl$ & $\phi_{\text{end}}/\mpl$ & $r$ \\
\hline
0.244301 & $5.55 \times 10^{-14}$ & -0.85224 & -0.19227 & $8.3 \times 10^{-4}$ \\
\hline
0.15     & $2.10 \times 10^{-14}$ & -0.59629 & -0.15263 & $3.1 \times 10^{-4}$ \\
\hline
0.0814338& $6.19 \times 10^{-15}$ & -0.37341 & -0.10725 & $9.3 \times 10^{-5}$ \\
\hline
0.06     & $3.36 \times 10^{-15}$ & -0.29344 & -0.08811 & $5.0 \times 10^{-5}$ \\
\hline
0.04     & $1.49 \times 10^{-15}$ & -0.21184 & -0.06682 & $2.2 \times 10^{-5}$ \\
\hline
0.035    & $1.14 \times 10^{-15}$ & -0.19003 & -0.06080 & $1.7 \times 10^{-5}$ \\
\hline
0.03     & $8.41 \times 10^{-16}$ & -0.16751 & -0.05442 & $1.3 \times 10^{-5}$ \\
\hline
0.028    & $7.33 \times 10^{-16}$ & -0.15827 & -0.05176 & $1.1 \times 10^{-5}$ \\
\hline
0.026    & $6.32 \times 10^{-16}$ & -0.14889 & -0.04902 & $9.4 \times 10^{-6}$ \\
\hline
0.025    & $5.84 \times 10^{-16}$ & -0.14415 & -0.04763 & $8.7 \times 10^{-6}$ \\
\hline
0.024    & $5.38 \times 10^{-16}$ & -0.13936 & -0.04621 & $8.0 \times 10^{-6}$ \\
\hline
0.022    & $4.52 \times 10^{-16}$ & -0.12966 & -0.04332 & $6.8 \times 10^{-6}$ \\
\hline
0.02     & $3.74 \times 10^{-16}$ & -0.11978 & -0.04033 & $5.6 \times 10^{-6}$ \\
\hline
0.018    & $3.03 \times 10^{-16}$ & -0.10970 & -0.03725 & $4.5 \times 10^{-6}$ \\
\hline
0.017    & $2.70 \times 10^{-16}$ & -0.10458 & -0.03566 & $4.0 \times 10^{-6}$ \\
\hline
0.0165   & $2.55 \times 10^{-16}$ & -0.10199 & -0.03486 & $3.8 \times 10^{-6}$ \\
\hline
0.016    & $2.39 \times 10^{-16}$ & -0.09939 & -0.03405 & $3.6 \times 10^{-6}$ \\
\hline
0.015    & $2.10 \times 10^{-16}$ & -0.09415 & -0.03240 & $3.1 \times 10^{-6}$ \\
\hline
0.014    & $1.83 \times 10^{-16}$ & -0.08884 & -0.03073 & $2.7 \times 10^{-6}$ \\
\hline
0.0125   & $1.46 \times 10^{-16}$ & -0.08074 & -0.02814 & $2.2 \times 10^{-6}$ \\
\hline
0.01     & $9.35 \times 10^{-17}$ & -0.06682 & -0.02363 & $1.4 \times 10^{-6}$ \\
\hline
\end{tabular}
\caption{Initial and end scalar field values for various $\mu$ characteristic scales, including tensor-to-scalar ratio $r$, as calculated at the pivot scale with $N_* = 60$ \efolds.}
\label{tab:phi-lambda-data}
\end{table}
\end{document}